\documentclass[aps,pra,epsfigure,twocolumn,superscriptaddress]{revtex4-2}
\usepackage[colorlinks=true,linkcolor=blue,urlcolor=blue,citecolor=blue,pdfusetitle]{hyperref}
\usepackage[utf8]{inputenc}
\usepackage[english]{babel}
\usepackage{amsmath}
\usepackage[caption = false]{subfig}
\usepackage{graphicx,epstopdf}
\usepackage{blindtext}
\usepackage[table,xcdraw]{xcolor}
\usepackage{lipsum}
\usepackage{amsfonts}
\usepackage{bbold}
\usepackage{bbm}
\usepackage{amssymb}
\usepackage{enumerate}
\usepackage{color}
\usepackage{xcolor}
\usepackage{latexsym}
\usepackage{physics}
\usepackage{braket}
\usepackage{times,txfonts}
\usepackage{bm}

\begin{document}

\title{Superradiant LIDAR}

\author{T. Kullick}
\affiliation{Friedrich-Alexander-Universität Erlangen-Nürnberg, Quantum Optics and Quantum Information, Staudtstr. 1, 91058 Erlangen, Germany}
\author{M. Bojer}
\affiliation{Friedrich-Alexander-Universität Erlangen-Nürnberg, Quantum Optics and Quantum Information, Staudtstr. 1, 91058 Erlangen, Germany}
\author{J. von Zanthier}
\email{joachim.vonzanthier@fau.de}
\affiliation{Friedrich-Alexander-Universität Erlangen-Nürnberg, Quantum Optics and Quantum Information, Staudtstr. 1, 91058 Erlangen, Germany}
\author{G. S. Agarwal}
\affiliation{Institute for Quantum Science and Engineering, Texas A\&M University, College Station, Texas 77843, USA}

\date{\today}

\begin{abstract}
In recent years, light detection and ranging (LIDAR) has seen a steep rise in the sensitivity of measuring the distances of remote objects. Here, we propose to enhance the sensitivity of LIDAR even further by exploiting Dicke's concept of superradiance, i.e., the  collective light emission of statistically independent light sources. By using $N$ thermal light sources (TLS) and measuring intensity correlations of order $m \geq 2$ instead of $m=1$, i.e., the intensity, we show that the Cramér-Rao bound on the measurement of the distance of a remote object undercuts that of traditional LIDAR by a factor of $N$, and can be reduced further with increasing correlation order $m$. Our numerical calculations
are supported by analytical expressions for the special cases of two and three TLS and a general approximate expression for any number of TLS.
\end{abstract}

\maketitle


\textit{Introduction.---}As a cornerstone of modern sensing techniques, light detection and ranging (LIDAR) plays a key role in a wide range of applications, from geospatial mapping~\cite{landslidesLIDAR_Jaboyedoff2012,  agricultureLIDAR_Rivera2023, SlopeInspectionLIDAR_Liu2025} and ecosystem characterization~\cite{maeda_2025, NPPwithLIDAR_Chen2025}, to autonomous vehicle navigation~\cite{autonomousVehicles_Yeong2021, autonomousVehicles_Song2024, autonomousVehicles_Ali2026}.
This has led to an increasingly active research field showcasing the rich potential of LIDAR~\cite{ thresholdQuantumLIDAR_Cohen2019, ultrafastLIDAR_Li2024, highResLongDistanceLIDAR_McCarthy2025, 4DLIDARimaging_Settembrini2026}. 
Beyond the widely employed incoherent LIDAR measuring the intensity of the retro-reflected light, interferometric coherent LIDAR setups make use of a phase-sensitive detection based on the interference between the incoming and a reference signal, yielding enhanced sensitivity and ambient-noise suppression~\cite{coherentLIDAR_Riemensberger2020, coherentLIDAR_Sambridge2021}. 
Whereas the maximum detection range of both LIDAR schemes is limited by the laser power, the range of coherent LIDAR systems is additionally constrained by the coherence length of the incoming light, with random phase noise deteriorating the signal quality~\cite{coherentLIDAR_rangeLimitation_Harris1998, coherentLIDAR_Sambridge2021, coherentLIDAR_Riemensberger2020, coherentLIDAR_rangeLimitation_Bayer2021}. 
Such interference-based systems will suffer in particular from atmospheric turbulence. 

$70$ years ago, Hanbury Brown and Twiss (HBT) proposed intensity interferometry for high resolution measurements of stellar diameters. 
The method uses intensity-intensity correlations rather than the measurement of the intensity itself and as such is insensitive to atmospheric turbulence. Hereby, according to the van Cittert-Zernike theorem, thermal light from stellar sources leads to nonzero spatial coherence at two telescopes, which governs the intensity-intensity correlation signal $G^{(2)}(r_1,r_2)$, 
the latter determining in turn the diameter of the star. Since the work of HBT, this method has been extensively used in many imaging applications, also beyond stellar interferometry~\cite{idi2017,2photonLIDAR_Tamma2023,quantumcoherenceLIDAR_Qian2023,intensityinterferometry_Liu2025}. 
Moreover, in many laboratory experiments, one generates thermal light from a laser source so that one has a high flux of photons besides obtaining a higher coherence length than that of a typical natural thermal source~\cite{Estes:71}.  Hence, intensity interferometry with laboratory produced thermal light continues to have great potential for many applications~\cite{quantumimaging_Shih2003, Shih_2005, HBTandTLSGhostImaging_Wang2009, GhostImaging_Padgett2017,ReviewGhotsImaging_Hoenders2018}.

In a recent work, Lee \textit{et al.}~\cite{2photonLIDAR_Tamma2023} introduced the concept of coherent two-photon LIDAR based on two-photon interference of incoherent light, exploiting the second-order correlation function $G^{(2)}(r_1,r_2)$ of two thermal light sources (TLS) (see also~\cite{Kurtsiefer_2023,staffas_2025}).
The authors showed experimentally that two-photon LIDAR bypasses the coherence-length-induced range limitation while being robust to atmospheric turbulence and ambient noise. Further, they demonstrated higher sensitivity in the measurement of the distance between a remote object and the light source. At the same time, Liu \textit{et al.}~\cite{intensityinterferometry_Liu2025} used laboratory produced thermal illumination and intensity interferometry to image mm-sized targets located 1.36\,km away and obtained resolution enhancement of about 14 times over the diffraction limit of a single telescope. We also mention that, besides the use of classical sources, quantum protocols for LIDAR have been discussed, based on the use of entangled light sources. For example, Reichert \textit{et al.}~\cite{heisenbergLIDAR_Reichert2024} have proposed Heisenberg limited quantum LIDAR, whereas Qian \textit{et al.}~\cite{quantumcoherenceLIDAR_Qian2023} have successfully demonstrated LIDAR based on quantum induced coherence using entangled photons from down converted sources~\cite{inducedcoherence_Zou1991}.

Apart from exploiting intensity correlations of second-order, we stress that one can also make use of higher-order intensity correlations, which considerably expands the landscape of applications of intensity interferometry~\cite{3Dsuperresolution_Dertinger2009,superresolutionfluorescence_Dertinger2012,superresolutionmicroscopy_Schwartz2013,noninvasive_Katz2014,3DSIM_Classen2018}. In that context, Oppel \textit{et al.}~\cite{magicPositionSetup_Oppel2012} demonstrated that the measurement of $m^{\mathrm{th}}$-order correlations produced by $m$ TLS can lead to higher resolution in the measurement of the distance between such sources.

\begin{figure}[t!]
    \includegraphics[width=\columnwidth]{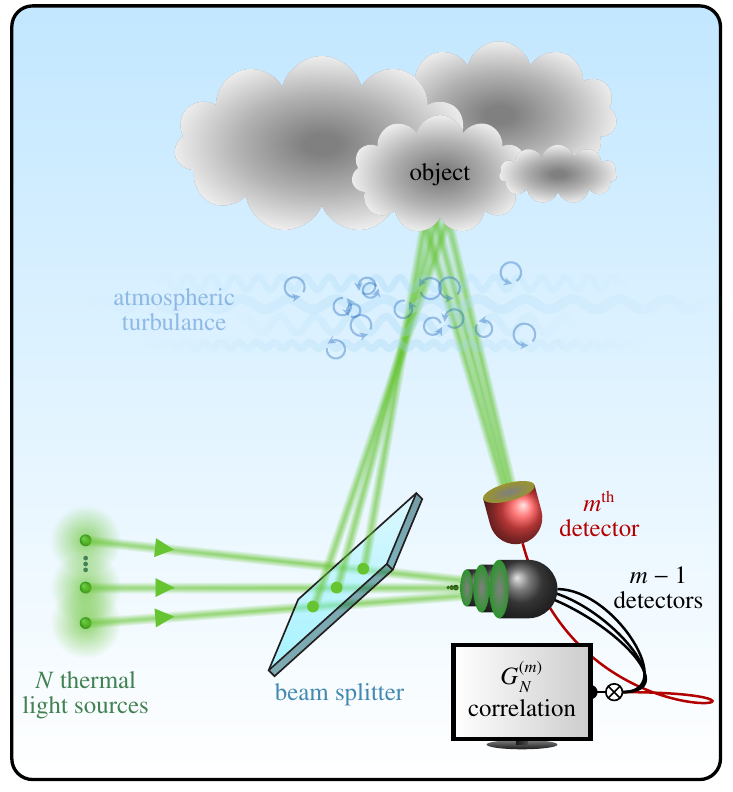}
    \caption{Sketch of a practical implementation of Superradiant LIDAR. $N$ equidistant TLS emit light that is separated by a beam splitter. The transmitted light is detected by $m-1$ detectors at a known distance. The reflected light propagates towards a remote object. After being reflected by the remote object, the light travels towards the $m^\mathrm{th}$ detector. Correlating the photons of all $m$ detectors yields a measurement of the $m^\mathrm{th}$-order correlation function $G^{(m)}_{N}(\delta_1, \delta_2) $, from which the unknown distance to the remote object can be extracted. Note that the measurement is insensitive to atmospheric turbulence.
    }
    \label{fig: application LIDAR}
\end{figure}

\begin{figure*}
    \centering
    \includegraphics[width = 0.9\textwidth]{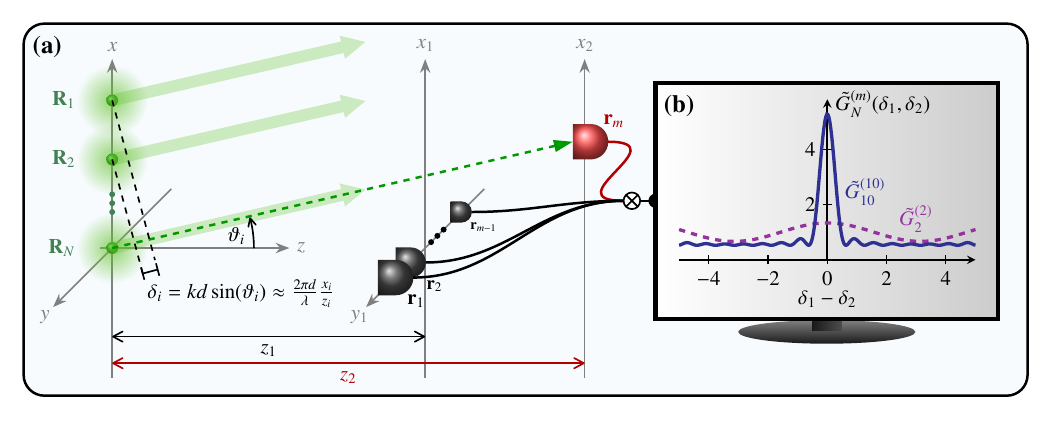}
    \caption{(a) Conceptual scheme of Superradiant LIDAR. $N$ equidistantly spaced, identical, statistically independent TLS are placed at positions $\mathbf{R}_l$, $l\in\{1, \dots, N\}$, along the $x$-axis. 
    The radiated light is detected by $m$ detectors at positions $\mathbf{r}_j$, $j\in\{1,\dots,m\}$, in the far field. The $m^\mathrm{th}$ detector is placed at an unknown distance $z_2$ under a variable angle $\vartheta_2 \approx  x_2 / z_2$, whereas the remaining $m-1$ detectors are placed at a distance $z_1$ under the angle $\vartheta_1 \approx  x_1/ z_1$.
    (b) The $m^\mathrm{th}$-order correlation function of $N$ TLS of the Superradiant LIDAR setup $\tilde{G}^{(m)}_{N}(\delta_1, \delta_2)$ is shown for the cases $N=m=2$ (purple) and $N=m=10$ (blue), highlighting that higher-order correlation functions of $N>2$ light sources yield steeper slopes.}
    \label{fig: Setup}
\end{figure*}

In this letter, we propose the idea of Superradiant LIDAR. It is derived from Dicke's concept of collective emission of spontaneous radiation from a system of atoms confined to dimensions much smaller than the wavelength prepared in particular states, known as superradiance~\cite{Dicke_1954}. In recent years, Dicke's ideas have been extended to larger systems~\cite{ magicPositionSetup_Oppel2012,Guerin2016,Asenjo2020,Wang2024,Robicheaux2024}. In particular, it has been shown that in this regime constructive quantum interference for special positions of detectors can mimic Dicke's superradiant emission behavior~\cite{simulatingsuperradiance_Wiegner2015,interference_Blatt2018,collective_Richter2023}. Already earlier, but inspired by the same ideas of constructive quantum interference and superradiance, Oppel \textit{et al.}~\cite{superradianceSetup_Oppel2014} showed that the measurement of higher-order correlation functions of statistically independent incoherent light sources improves the angular resolution by the measured correlation order. We here propose a setup that realizes a LIDAR scheme by use of higher-order correlation functions (see Figs.~\ref{fig: application LIDAR} and~\ref{fig: Setup}). In particular, we analyze its sensitivity by calculating the corresponding Cramér-Rao bound. Our findings demonstrate that exploiting higher-order correlation functions in a LIDAR setup results in a larger sensitivity in the determination of the distance between source and remote object than that for two-photon LIDAR. Hereby, similar to HBT, Superradiant LIDAR would be immune to atmospheric turbulence and ambient noise.


\textit{Concept of Superradiant LIDAR.---}
The scheme of our Superradiant LIDAR setup is depicted in Fig.~\ref{fig: Setup}(a).
Similar to~\cite{magicPositionSetup_Oppel2012, superradianceSetup_Oppel2014, magicPositionSetup_Cao2015, superradianceSetup_Bhati2016}, we consider $N$ identical statistically independent TLS emitting light of wavelength $\lambda$ at equidistant positions $\mathbf{R}_l$ with $l\in\{1,\dots,N\}$ along the $x$-axis.
The radiated thermal light is detected coincidentally in the far field by $m$ detectors at positions $\mathbf{r}_j$ with $j\in\{1,\dots,m\}$. This amounts to measuring the $m^\mathrm{th}$-order photon correlation function~\cite{Glauber_1963}
\begin{equation}
	G^{(m)}_N(\{\mathbf{r}_j\})=\braket{\hat{E}^{(-)}(\mathbf{r}_1)\dots\hat{E}^{(-)}(\mathbf{r}_m)\hat{E}^{(+)}(\mathbf{r}_m)\dots\hat{E}^{(+)}(\mathbf{r}_1)}_{\hat{\rho}},
	\label{eq:corr_func}
\end{equation}
where
\begin{equation}
	\label{eq:Epos}
	\hat{E}^{(+)}(\mathbf{r}_j) = \sum_{l=1}^{N} e^{-i k \hat{\mathbf{r}}_j \mathbf{R}_l} \hat{a}_l
\end{equation}
is the positive frequency part of the electric field operator evaluated at position $\mathbf{r}_j$, with $\hat{\mathbf r}_j$ being a unit vector in the direction of observation and $k = 2\pi/\lambda$ the wave number.
Here, $\hat{a}_l$ is the photon annihilation operator of the $l^\mathrm{th}$ TLS, while $\hat{\rho}$ denotes the density matrix of the light field produced by the $N$ TLS~\cite{superradianceSetup_Bhati2016}.
Note that in Eq.~\eqref{eq:Epos}, we set the electric field amplitude to unity and used a scalar notation since we consider only a single polarization of the field emitted by the TLS.

In the following, we restrict the positions of $m-1$ detectors to the same angle $\vartheta_1$ with corresponding phase difference $\delta_1=kd \sin(\vartheta_1)$ between adjacent sources, realized by stacking the detectors along the third dimension ($y$-axis), while the $m^\mathrm{th}$ detector is placed under a variable angle $\vartheta_2$ and phase difference $\delta_2= kd \sin(\vartheta_2)$ (see Fig.~\ref{fig: Setup}(a)).
Considering measurements in the far field, the phase differences $\delta_i$, $i\in\{1,2\}$, can be expressed via the positions of the detectors $x_i$ and their distance to the light source $z_i$ as $\delta_i = k d\sin(\vartheta_i) \approx 2\pi dx_i/(\lambda z_i)$.

We note that up to now our setup coincides with the generalized HBT setup implemented by Oppel \textit{et al.}~\cite{superradianceSetup_Oppel2014}.
Yet, whereas Oppel \textit{et al.} showed that directional superradiant emission can be realized with thermal light via measurements of higher-order correlation functions, we propose here a modified scheme to precisely determine the unknown distance $z_2$.
Namely, we place the $m^\mathrm{th}$ detector at a different distance $z_2$ from the light sources than the remaining detectors, which are located at a distance $z_1$, see Fig.~\ref{fig: Setup}(a).
Investigating the $m^{\text{th}}$-order correlation function for this setup, we obtain~\cite{superradianceSetup_Oppel2014,superradianceSetup_Bhati2016}
\begin{equation}
    \label{eq: correlation function superradiance}
    \Tilde{G}^{(m)}_{N}(\delta_1, \delta_2) 
    = \left(1+\frac{m-1}{N}\right)^{-1}
    \left(1+\frac{m-1}{N^2}\frac{\sin^2{\left[N\frac{\delta_1-\delta_2}{2}\right]}}{\sin^2{\left[\frac{\delta_1-\delta_2}{2}\right]}}\right),
\end{equation}
where we normalized $G^{(m)}_{N}(\delta_1, \delta_2)$ by the spatial average of the correlation function (see Supplemental Material~\cite{SupplementalMaterial}).
As illustrated in Fig.~\ref{fig: Setup}(b), 
$\Tilde{G}^{(m)}_{N}(\delta_1, \delta_2)$ displays an enhanced directional radiation pattern with steep slopes for $N>2$ TLS.
Intuitively, we would expect that this results in a higher Fisher information and thus increased sensitivity in measuring the distance $z_2$.

A simple setup that realizes our Superradiant LIDAR scheme in the laboratory is proposed in~\cite{SupplementalMaterial}.
Instead of employing $m$ single-photon detectors, we could correlate $m$ pixels from two digital charge-coupled device cameras (CCD~1 and CCD~2), thereby taking advantage of a significantly larger amount of data provided by their high pixel count~\cite{magicPositionSetup_Oppel2012, superradianceSetup_Oppel2014}.
In particular, we can take $m-1$ subsequent pixels along the $y_1$-axis of CCD~1 under a fixed angle $\vartheta_1 \propto x_1 = pn_1$ and correlate them with one pixel varying along $\vartheta_2 \propto x_2 = pn_2$ on CCD 2, yielding a measurement of $\Tilde{G}^{(m)}_N(n_1,n_2)$.
Here, we use the fact that every detector position on the CCD can be expressed as its pixel width $p$ times the corresponding pixel number $n_{i}$, $i\in\{1,2\}$.
In this case the phase differences become $\delta_i = \omega_i n_i$, with $\omega_i = 2 \pi dp/(\lambda z_i)$ being the corresponding spatial frequency.
This procedure is repeated for every column of CCD 1 (every $n_1$), see~\cite{SupplementalMaterial}.
Considering $N_\mathrm{H}$ horizontal pixels, we obtain a set of $N_\mathrm{H}$ $m^\mathrm{th}$-order correlation functions, from which the unknown distance $z_2$ can be extracted via a fit (or a Fast Fourier Transform).
Note that our proposed Superradiant LIDAR setup reduces to the two-photon LIDAR setup reported in~\cite{2photonLIDAR_Tamma2023} in the special case of $N=m=2$.


\textit{Cramér-Rao bound of Superradiant LIDAR.---}Analyzing the Cramér-Rao bound of Superradiant LIDAR requires the calculation of the Fisher information.
The latter is entirely determined by the correlation function Eq.~\eqref{eq: correlation function superradiance}, assuming independent Poisson-distributed measurements~\cite{parameterEstimation_Li2017}. In this way, we obtain for the Cramér-Rao bound of Superradiant LIDAR
\begin{widetext}
\begin{align}
    \mathrm{Var}[\hat{z}_2]^{-1} \leq \mathcal{F}_{z_2}^{\text{(sup)}}(N,m)
    &=
    \label{eq: general Fisher information}
    \sum_{n_1 = 1}^{N_{\text{H}}}\sum_{n_2 = 1}^{N_{\text{H}}}\frac{1}{\Tilde{G}^{(m)}_{N}(n_1, n_2)}\left(\frac{\partial \Tilde{G}^{(m)}_{N}(n_1, n_2)}{\partial z_2}\right)^2 \\
    &\approx
    \label{eq: general Fisher information superradiance setup}
    \frac{\omega_2^2 N_{\text{H}}^4}{3z_2^2}\frac{1}{\pi}\left(1+\frac{m-1}{N}\right)^{-1}\left(\frac{m-1}{N^2}\right)^2 \int_{0}^{\pi}\mathrm{d}s \, \frac{\left(N\cos(Ns)\sin(Ns)\sin(s) - \sin^2(Ns)\cos(s)\right)^2}{\left(\sin^6(s)+\frac{m-1}{N^2}\sin^2(Ns)\sin^4(s)\right)},
\end{align}
\end{widetext}
where we assumed a sufficiently large number of horizontal pixels $N_{\text{H}}$ to approximate the sums by integrals and $\omega_1N_{\text{H}} \approx 2\pi\mathcal{N}$ with $\mathcal{N} \in \mathbb{N}$ (see~\cite{SupplementalMaterial} for details).

The numerical evaluation of this expression for $m\in\{2,3,\dots,20\}$ and $N\in\{2,3,\dots,20\}$ reveals the following behavior (see Fig.~\ref{fig: Fisher information superradiant LIDAR}):
\begin{figure}[h!]
    \includegraphics[width=\columnwidth]{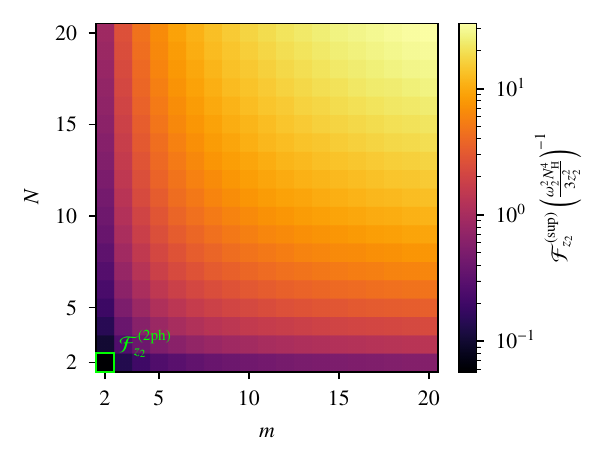}
    \caption{Fisher information $\mathcal{F}_{z_2}^{\text{(sup)}}$ of Superradiant LIDAR as a function of the correlation order $m\in\{2,3,\dots,20\}$ and the number of TLS $N\in\{2,3,\dots,20\}$.
    The special case $N=m=2$ (green rectangle), equivalent to two-photon LIDAR, marks the minimum of $\mathcal{F}_{z_2}^{\text{(sup)}}$.
    }
    \label{fig: Fisher information superradiant LIDAR}
\end{figure}
First of all, as mentioned above, the Fisher information of Superradiant LIDAR coincides with the one of two-photon LIDAR $\mathcal{F}^\mathrm{(2ph)}_{z_2}$ for the special case $N=m=2$ (green rectangle).
Yet, increasing the number of TLS $N$ and/or the order of correlation $m$ leads to an improvement of the Fisher information by up to almost 3 orders of magnitude in the displayed regime.
Hence, the Cramér-Rao bound of Superradiant LIDAR is smaller than that of two-photon LIDAR for any configuration except $N=m=2$.

Beyond numerical evaluations of the Fisher information, we also derive an analytical expression for two and three TLS.
Starting with $N=2$ TLS and  $m\geq 2$, the normalized correlation function reduces to $\Tilde{G}^{(m)}_{2}(\delta_1, \delta_2) = \left(1 + V\cos[\delta_1 - \delta_2]\right)$ with $V=(m-1)/(m+1)$ being the visibility of the $m^\mathrm{th}$-order correlation function~\cite{superradianceSetup_Oppel2014}.
It is evident that this correlation function exhibits the same functional dependence as in the case of two-photon LIDAR~\cite{2photonLIDAR_Tamma2023}.
We therefore conclude that the Fisher information is given by
\begin{equation}
    \label{eq: superradiant LIDAR's Fisher information for N=2}
    \mathcal{F}_{z_2}^{\text{(sup)}}(2,m) 
    =
    \frac{\omega_2^2 N_{\text{H}}^4}{3z_2^2}\left(1-\frac{2\sqrt{m}}{m+1}\right),
\end{equation}
which coincides with the numerical calculations.
A careful analysis of $N=3$ TLS reveals that the corresponding Fisher information is given  by
\begin{align}
    \label{eq: superradiant LIDAR's Fisher information for N=3}
    \begin{split}
    \mathcal{F}_{z_2}^{\text{(sup)}}(3,m)
    &=
    \frac{\omega_2^2 N_{\text{H}}^4}{3z_2^2}\frac{2}{3(m+2)} \\
    &\times\left(4m + 14 - 3\sqrt{6\left(m+2+\sqrt{m(m+8)}\right)}\right),
    \end{split}
\end{align}
highlighting the increased complexity of analytical solutions with an increasing number of TLS.


\textit{Lower Bound of the Fisher Information.---}Although no general analytical expression as a function of $N$ and $m$ for $N>3$ could be identified, we find the following approximate solution of the general Fisher information for arbitrary $N$ and $m$~\cite{SupplementalMaterial}
\begin{equation}
    \label{eq: Approximative Fisher information for superradiant LIADR}
    \mathcal{F}_{z_2}^{\text{(sup)}}(N,m)
    \approx
    \frac{\omega_2^2 N_{\text{H}}^4}{3z_2^2}\frac{N^2}{4(N+m-1)}
    \left(m + 1 - 2\sqrt{m}\right).
\end{equation}
This expression is constructed such that it reproduces the analytical solution of two TLS in Eq.~\eqref{eq: superradiant LIDAR's Fisher information for N=2} and inherits a linear increase in $N$ if the normalization prefactor is neglected; the latter is motivated from a physical point of view, as the angular width of the central maximum of the correlation function decreases proportionally to $N^{-1}$~\cite{superradianceSetup_Bhati2016}.
Furthermore, it serves as a lower bound of the Fisher information in general.

In order to validate the accuracy and underestimation of this expression, we calculated the relative difference $\Delta\mathcal{F}_\mathrm{rel} = (\mathcal{F}_\mathrm{num} - \mathcal{F}_\mathrm{app})/\mathcal{F}_\mathrm{num}$ of the numerical evaluation of Eq.~\eqref{eq: general Fisher information superradiance setup} and the approximate result given in Eq.~\eqref{eq: Approximative Fisher information for superradiant LIADR}.
As shown in~\cite{SupplementalMaterial}, for $N>3$, all deviations lie below $9\,\%$ in the considered regime. 
Moreover, we observe that the relative difference shrinks as the number of light sources $N$ increases.
This reveals that the physically motivated linear increase in $N$ dominates for large $N$ and models the Fisher information adequately.
We finally emphasize that the approximate expression of the Fisher information in Eq.~\eqref{eq: Approximative Fisher information for superradiant LIADR} always underestimates the actual value.
Hence, the approximate Fisher information for Superradiant LIDAR in Eq.~\eqref{eq: Approximative Fisher information for superradiant LIADR} serves as a sufficient lower bound to estimate the Cramér-Rao bound.


\textit{Conclusion.---}We introduced the concept of Superradiant LIDAR based on simulating Dicke's collective emission of spontaneous radiation by multiphoton interferences with $N$ independent TLS using higher-order correlation functions~\cite{superradianceSetup_Oppel2014}.
We numerically demonstrated that measuring higher-order correlation functions within the Superradiant LIDAR scheme leads to a  Fisher information improved by at least a factor of $N$ compared to that of two-photon LIDAR.
Hence, the Cramér-Rao bound decreases at least by the same amount, demonstrating that Superradiant LIDAR is profoundly more sensitive to distance measurements of remote objects than two-photon LIDAR. The numerical calculations are supported by analytical solutions for the special cases of two and three TLS and a general approximate expression that slightly overestimates the true Cramér-Rao bound.

Besides proposing an enhanced LIDAR setup, this result showcases how fundamental quantum effects such as Dicke superradiance can improve conventional measurement techniques known from classical optics.
It might lay the groundwork for a wide variety of other innovative quantum-inspired applications.


\textit{Acknowledgments.---}GSA thanks support from NSF award no. 2426699 and DOE, Fusion Energy Science, award no. DE-SC 0024882, IFE-STAR. JvZ and MB acknowledge funding by the Deutsche Forschungsgemeinschaft (DFG, German Research Foundation) - Project-ID 429529648 - TRR 306 QuCoLiMa (‘Quantum Cooperativity of Light and Matter’).

\bibliography{refs}


\clearpage
\onecolumngrid
\appendix


\onecolumngrid


\begin{center}
{\large{ {\bf Supplemental Material for: Superradiant LIDAR }}}

\setcounter{page}{1}

\vskip0.5\baselineskip{T. Kullick,$^{1}$ M. Bojer,$^{1}$ J. von Zanthier,$^{1}$ and G. S. Agarwal$^{2}$}

\vskip0.5\baselineskip{ {\it $^{1}$Friedrich-Alexander-Universität Erlangen-Nürnberg, Quantum Optics and Quantum Information, Staudtstr. 1, 91058 Erlangen, Germany\\
$^{2}$Institute for Quantum Science and Engineering, Texas A\&M University, College Station, Texas 77843, USA}}
\end{center}

\appendix

\setcounter{figure}{0}

\section{Considered Correlation Functions and their Normalization}
A well-defined and unified normalization of the considered correlation functions is essential for a proper comparison of the different setups. 
Hence, this appendix introduces the considered correlation functions and the normalizations chosen to compare our proposed Superradiant LIDAR setup with the coherent two-photon LIDAR setup by Lee \textit{et al.}~\cite{2photonLIDAR_Tamma2023}.

The theoretical considerations of two-photon LIDAR with its analytical visibility of $V=1/3$ are based on the correlation function~\cite{2photonLIDAR_Tamma2023}
\begin{equation}
    G^{(2)}_\mathrm{2ph}(\delta_1, \delta_2) = \Bar{I}^2\left(1 + V\cos\left[\delta_1 - \delta_2\right]\right).
\end{equation}
Thereby, $\Bar{I}$ denotes the mean pixel intensity of the experimental setup, and the phase differences $\delta_{i}$ are given by $\delta_{i} = \omega_{i}n_{i}$, with $\omega_{i}$ being the spatial frequency and $n_{i}$ denoting the pixel number (with $i\in\{1,2\}$).
On the other hand, the general $m^{\mathrm{th}}$-order correlation function of the superradiance setup of $N$ identical, uncorrelated thermal light sources is given by~\cite{superradianceSetup_Bhati2016}
\begin{equation}
    G^{(m)}_{N}(\delta_1, \delta_2) = (N\bar{n})^m(m-1)!\left(1+\frac{m-1}{N^2}\frac{\sin^2{\left[N\frac{\delta_1-\delta_2}{2}\right]}}{\sin^2{\left[\frac{\delta_1-\delta_2}{2}\right]}}\right),
\end{equation}
where we refer to the mean photon number per source as $\bar{n}$.
A comparison of the two correlation functions highlights the necessity of a proper normalization:
Whereas Lee \textit{et al.} based their analytical calculations on an experimentally motivated mean pixel intensity $\bar{I}$, we instead describe the correlation function with respect to the source intensity.
Hence, a normalization that properly scales the functional dependencies of the correlation functions is necessary.

We choose to normalize both correlation functions by integrating the correlation function over whole periods of $2\pi\mathcal{M}$ with $\mathcal{M}\in\mathbb{N}$, dividing by the length of $2\pi\mathcal{M}$, and demanding that value to be 1.
This corresponds to taking the spatial average of the correlation functions.
The correlation functions of the two-photon LIDAR and Superradiant LIDAR depend on two phase differences.
We fix one of them and set it to 0 for simplicity.
This leads to the following normalization factors:
\begin{itemize}
    \item[(i)] two-photon LIDAR
    \begin{equation}
        \frac{1}{2\pi\mathcal{M}}\int_0^{2\pi\mathcal{M}}\mathrm{d}\delta_1 \, G_{\mathrm{2ph}}^{(2)}
        =
        \frac{1}{2\pi\mathcal{M}}\int_0^{2\pi\mathcal{M}}\mathrm{d}\delta_1 \, \Bar{I}^2\left(1+\frac{1}{3}\cos[\delta_1]\right) = \Bar{I}^2
    \end{equation}

    \item[(ii)] Superradiant LIDAR
    \begin{equation}
        \frac{1}{2\pi\mathcal{M}}\int_0^{2\pi\mathcal{M}}\mathrm{d}\delta_1 \, G_{N}^{(m)} =
        \frac{1}{2\pi\mathcal{M}}\int_0^{2\pi\mathcal{M}}\mathrm{d}\delta_1 \, (N\bar{n})^m(m-1)!\left(1+\frac{m-1}{N^2}\frac{\sin^2[N\delta_1/2]}{\sin^2[\delta_1/2]}\right) = (N\bar{n})^m(m-1)!\left(1+\frac{m-1}{N}\right).
    \end{equation}
\end{itemize}
Evidently, the prefactors of the functional dependencies cancel with the introduced normalization.
Moreover, the correlation function of Superradiant LIDAR reduces to that of two-photon LIDAR for the case $N=m=2$.
This is important, as both setups are identical in this special case.
Hence, we conclude that the above-introduced normalization is suitable for our theoretical investigations of the Fisher information.
We denote the normalized correlation function by $\Tilde{G}^{(m)}_N$.

\section{Detailed Description of the Superradiant LIDAR Setup}
\begin{figure*}
    \centering
    \includegraphics[width = 0.9\textwidth]{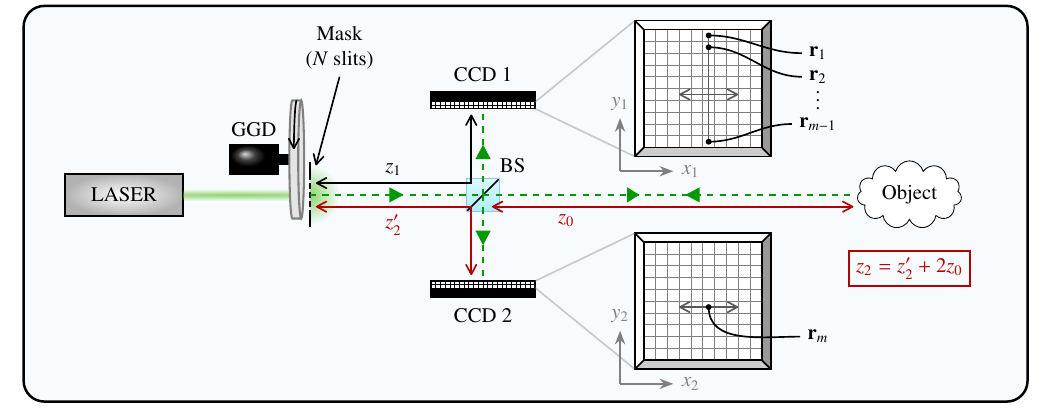}
    \caption{Detailed proposal for a Superradiant LIDAR setup. For details, see text. GGD: rotating ground-glass disk, BS: beam splitter, CCD: charge-coupled device camera.}
    \label{fig app: detailed Setup}
\end{figure*}

In order to illustrate the simplicity of Superradiant LIDAR,
Fig.~\ref{fig app: detailed Setup} shows a detailed scheme that emulates the Superradiant LIDAR setup.
$N$ uncorrelated thermal light sources are realized by directing a laser beam onto a rotating ground-glass disk (GGD) and placing a mask with $N$ equidistantly spaced slits behind it~\cite{magicPositionSetup_Oppel2012,superradianceSetup_Oppel2014}.
The emerging beam of the generated thermal light is split by a beam splitter (BS).
The reflected component is detected by a charge-coupled device camera (CCD~1) along the known path $z_1$.
The transmitted beam propagates towards a mirror that simulates a remote object, and the light reflected from the mirror (remote object) is subsequently redirected by the BS and detected by a second camera (CCD~2) at an unknown distance $z_2 = z_2' + 2z_0$.

As shown in Fig.~\ref{fig app: detailed Setup} and described in the main manuscript, we correlate $m$ pixels from two digital cameras instead of using $m$ single-photon detectors.
The positions $\mathbf{r}_j$ with $j\in\{1,2,\dots,m\}$ of the detectors shown in Fig.~2(a) in the main script are also displayed in Fig.~\ref{fig app: detailed Setup} to illustrate the one-to-one correspondence between the two visualizations.
It is evident that our proposed Superradiant LIDAR scheme reduces to the two-photon LIDAR setup in the special case of $N=m=2$ \cite{2photonLIDAR_Tamma2023}.
The measurement procedure to obtain a set of $N_\mathrm{H}$ $m^\mathrm{th}$-order correlation functions and to extract the unknown distance $z_2$ is thoroughly explained in the main script.

\section{Calculation of the Numerically Evaluated Expression of the Fisher Information}
This appendix details the derivation of the Superradiant LIDAR's Fisher information (Eq.~(5) in the main text) that is numerically evaluated.
We start by rewriting the normalized correlation function $\Tilde{G}^{(m)}_N(\delta_1, \delta_2)$ in terms of the pixel numbers $n_i$ (with $i\in\{1,2\}$) by approximating the phase differences via $\delta_i \approx \omega_in_i$ with $\omega_i = 2\pi d p/(\lambda z_i)$.
Thereby, $p$ is the pixel width and $d$ denotes the equal distance between the identical thermal light sources of wavelength $\lambda$ that are located at distance $z_i$ from the corresponding detectors.
Furthermore, it is convenient to introduce the notation $S_X \equiv \sin[X(\omega_1n_1 - \omega_2n_2)/2]$ and $C_X \equiv \cos[X(\omega_1n_1 - \omega_2n_2)/2]$.

With the assumption of Poisson-distributed measurements and the proposed Superradiant LIDAR setup (see Fig.~\ref{fig app: detailed Setup} in the SM or Fig.~2 in the main text), the Fisher information for the unknown distance $z_2$ can be derived via~\cite{parameterEstimation_Li2017}
\begin{equation}
    \label{eq app: general Fisher information}
    \mathcal{F}_{z_2}^{\text{(sup)}}(N,m) =
    \sum_{n_1 = 1}^{N_{\text{H}}}\sum_{n_2 = 1}^{N_{\text{H}}}\frac{1}{\Tilde{G}^{(m)}_{N}(n_1, n_2)}\left(\frac{\partial \Tilde{G}^{(m)}_{N}(n_1, n_2)}{\partial z_2}\right)^2,
\end{equation}
where $N_\mathrm{H}$ denotes the number of horizontal pixels of the camera.
Determining the derivative of the correlation function with respect to $z_2$ and inserting it in Eq.~\eqref{eq app: general Fisher information} leads to
\begin{align}
    \mathcal{F}^{\text{(sup)}}_{z_2}(N,m)
    &=
    \frac{\omega_2^2}{z_2^2}\left(1+\frac{m-1}{N}\right)^{-1}\left(\frac{m-1}{N^2}\right)^2\sum_{n_1 = 1}^{N_\text{H}}\sum_{n_2 = 1}^{N_\text{H}}n_2^2\frac{\left(NC_NS_NS_1-C_1S_N^2\right)^2}{\left(1+\frac{m-1}{N^2}\frac{S_N^2}{S_1^2}\right)S_1^6} \\
    &\approx
    \frac{\omega_2^2}{z_2^2}\left(1+\frac{m-1}{N}\right)^{-1}\left(\frac{m-1}{N^2}\right)^2\int_{\left[0,N_\text{H}\right]}\mathrm{d}^2n \, n_2^2\frac{\left(NC_NS_NS_1-C_1S_N^2\right)^2}{\left(S_1^6+\frac{m-1}{N^2}S_N^2S_1^4\right)},
\end{align}
where we assumed a sufficiently large number of horizontal pixels $N_\mathrm{H}$, such that we can approximate the sums by integrals.

By substituting $s = (\omega_1n_1-\omega_2n_2)/2$, and rewriting the trigonometric notations as $S_X \equiv \sin[Xs]$ and $C_X \equiv\cos[Xs]$, we obtain
\begin{equation}
    \mathcal{F}^{\text{(sup)}}_{z_2}(N,m)
    \approx
    \frac{\omega_2^2}{z_2^2}\left(1+\frac{m-1}{N}\right)^{-1}\left(\frac{m-1}{N^2}\right)^2\frac{2}{\omega_1}
    \int_{0}^{N_\text{H}}\mathrm{d}n_2 \, n_2^2
    \int_{0}^{\frac{\omega_1 N_\text{H}}{2}}\mathrm{d}s\,\frac{\left(NC_NS_NS_1-C_1S_N^2\right)^2}{\left(S_1^6+\frac{m-1}{N^2}S_N^2S_1^4\right)}.
\end{equation}
Note that the integration boundaries are $[0,\omega_1N_\text{H}/2]$ instead of $[-\omega_2n_2/2,(\omega_1N_\text{H}-\omega_2n_2)/2]$.
This is justified as long as we are integrating over whole periods of the $\pi$-periodic trigonometric integrand.
Therefore, the integral over $\mathrm{d}n_2$ becomes independent of the second integral and can be easily integrated.

In particular, we assume $\omega_1N_\mathrm{H}\approx2\pi\mathcal{N}$ with $\mathcal{N}\in\mathbb{N}$, which satisfies the latter imposed condition.
Moreover, it is in accordance with the analytical derivation of two-photon LIDAR's Fisher information~\cite{2photonLIDAR_Tamma2023}, which is crucial as Superradiant LIDAR recovers two-photon LIDAR in the special case of $N=m=2$.

Assuming $\omega_1N_\mathrm{H}\approx2\pi\mathcal{N}$ leads together with the $\pi$-periodicity of the integrand to
\begin{align}
    \mathcal{F}^{\text{(sup)}}_{z_2}(N,m)
    &\approx
    \frac{\omega_2^2}{z_2^2}\left(1+\frac{m-1}{N}\right)^{-1}\left(\frac{m-1}{N^2}\right)^2\frac{2N_\text{H}^3}{3\omega_1}
    \int_{0}^{\pi\mathcal{N}}\mathrm{d}s \, \frac{\left(NC_NS_NS_1-C_1S_N^2\right)^2}{\left(S_1^6+\frac{m-1}{N^2}S_N^2S_1^4\right)} \\
    &=
    \frac{\omega_2^2}{z_2^2}\left(1+\frac{m-1}{N}\right)^{-1}\left(\frac{m-1}{N^2}\right)^2\frac{N_\text{H}^3}{3\pi}\frac{2\pi\mathcal{N}}{\omega_1}
    \int_{0}^{\pi}\mathrm{d}s \, \frac{\left(NC_NS_NS_1-C_1S_N^2\right)^2}{\left(S_1^6+\frac{m-1}{N^2}S_N^2S_1^4\right)} \\
    &=
    \label{eq app: superradiance numerical integral}
    \frac{\omega_2^2 N_{\text{H}}^4}{3z_2^2}\frac{1}{\pi}\left(1+\frac{m-1}{N}\right)^{-1}\left(\frac{m-1}{N^2}\right)^2 \int_{0}^{\pi}\mathrm{d}s \, \frac{\left(N\cos(Ns)\sin(Ns)\sin(s) - \sin^2(Ns)\cos(s)\right)^2}{\left(\sin^6(s)+\frac{m-1}{N^2}\sin^2(Ns)\sin^4(s)\right)}.
\end{align}
Thus, we finally arrive at Eq.~(5) of the main text, which is the expression that is numerically evaluated.

\section{Motivation of a Lower Bound and Fit Function Approximation of the Fisher Information of Superradiant LIDAR}
As can be seen, the Superradiant LIDAR's Fisher information for $N=2,3$ light sources (Eqs.~(6) and~(7) in the main text) have a similar shape.
Moreover, all numerical solutions show a similar behavior.
Hence, we exploit this similarity to obtain a general lower bound on the Fisher information and additionally model it using a fit.
The motivation of the lower-bound expression and of the fit function are subject of this appendix.
Additionally, we evaluate some fit results.

It is a good starting point to investigate Superradiant LIDAR's Fisher information for three thermal light sources in the limits of small and large orders of $m$.
We are considering the Fisher information without the prefactor $\omega_2^2 N_{\text{H}}^4/(3z_2^2\left(1+(m-1)/N\right))$, as the functional dependence on $N$ and $m$ is exactly known for this term.
We introduce this quantity as $F_{z_2}^{\text{(sup)}}(N,m)$.

For small correlation orders $m\ll1$, we get
\begin{align}
    \label{eq app: unnormalized Fisher information N=3 for large m}
    F_{z_2}^{\text{(sup)}}(3,m)
    \overset{m\ll1}{\approx}
    \left(4m+\left(14-6\sqrt{3}\right)-3\sqrt{6}\sqrt{m}\right),
\end{align}
by performing a Taylor expansion at $m=0$ to leading order.
For large correlation orders $m\gg1$, we can approximate the nested square root of the analytical solution as
\begin{align}
    F_{z_2}^{\text{(sup)}}(3,m)
    \overset{m\gg1}{\approx}
    \frac{2}{3^2}\left(4m+14-3\sqrt{12}\sqrt{m}\right).
\end{align}
Comparing the approximated forms of $F_{z_2}^{\text{(sup)}}(3,m)$ with the Fisher information of two thermal light sources, rewritten as 
\begin{equation}
    \label{eq app: unnormalized Fisher information N=2}
    F_{z_2}^{\text{(sup)}}(2,m)
    =
    \frac{1}{2}\left(m+1-2\sqrt{m}\right),
\end{equation}
reveals an evident functional similarity in $m$.
These findings suggest that it may be possible to construct a lower bound for the general Fisher information based on the analytical solution of two thermal light sources.
This can be done by using its $m$-dependence $(m+1-2\sqrt{m})$ as the lower bound and introducing an $N$-dependent growth because the coefficients increase in both $m$-limits of $F_{z_2}^{\text{(sup)}}(3,m)$.
As the angular width of the central maximum of the correlation function decreases proportional to $N^{-1}$~\cite{superradianceSetup_Bhati2016}, and because the Fisher information depends on the slope of the correlation function, we introduce a linear growth in $N$ so that it reproduces the analytical solution of two thermal light sources in Eq.~\eqref{eq app: unnormalized Fisher information N=2}.
Hence, we approximate the general unnormalized Fisher information by
\begin{equation}
    \label{eq app: unnormalized lower bound Fisher information}
    F_{z_2}^{\text{(sup)}}(N,m)
    \approx
    \frac{N}{4}\left(m+1-2\sqrt{m}\right).
\end{equation}
Multiplying Eq.~\eqref{eq app: unnormalized lower bound Fisher information} by the above neglected prefactor leads to
\begin{equation}
    \label{eq app: Approximative Fisher information for superradiant LIADR}
    \mathcal{F}_{z_2}^{\text{(sup)}}(N,m)
    \approx
    \frac{\omega_2^2 N_{\text{H}}^4}{3z_2^2}\frac{N^2}{4(N+m-1)}
    \left(m + 1 - 2\sqrt{m}\right),
\end{equation}
which corresponds to Eq.~(8) in the main manuscript.

\begin{figure}[h!]
    \includegraphics[scale = 0.9]{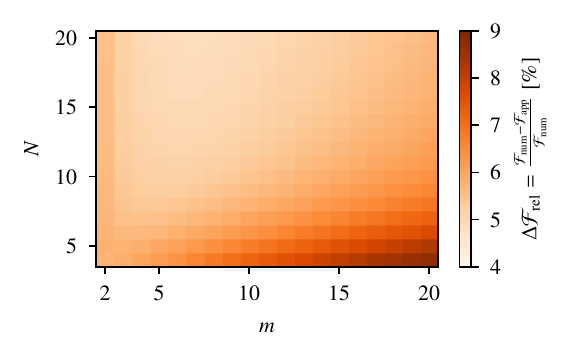}
    \caption{The relative difference $\Delta\mathcal{F}_\mathrm{rel}$ between the numerically evaluated Fisher information of Superradiant LIDAR (Eq.~\eqref{eq app: superradiance numerical integral}) and the approximated Fisher information in Eq.~\eqref{eq app: Approximative Fisher information for superradiant LIADR} displayed for $N\in\{4,5,\dots,20\}$ and $m\in\{2,3,\dots,20\}$.
    The cases $N=2,3$ are not shown, as analytical solutions are presented in Eqs.~(6) and~(7) in the main text, respectively.
    All deviations are below $9\,\%$, and the approximate form always underestimates the numerical results.}
    \label{fig app: lower bound discussion}
\end{figure}

Fig.~\ref{fig app: lower bound discussion} displays the relative difference $\Delta\mathcal{F}_\mathrm{rel} = (\mathcal{F}_\mathrm{num} - \mathcal{F}_\mathrm{app})/\mathcal{F}_\mathrm{num}$ of the numerical evaluation of Eq.~\eqref{eq app: superradiance numerical integral}  and the approximate result in Eq.~\eqref{eq app: Approximative Fisher information for superradiant LIADR} to verify the accuracy of the latter equation.
The discussion in the main text shows that this approximate expression of the Fisher information adequately models the Fisher information (all deviations are below 9\,\%) and serves as a lower bound, as it always underestimates the true value.

However, one can find a more precise approximation of the Fisher information by using a fit function as we show in the following.
Motivated by Eqs.~\eqref{eq app: unnormalized Fisher information N=3 for large m} and~\eqref{eq app: unnormalized Fisher information N=2}, we generalize the similarity by defining the fit function as
\begin{equation}
    \label{eq app: fit Fisher information}
    \mathcal{F}_{z_2}^{\text{(fit)}}(N,m)
    =
    \frac{\omega_2^2 N_{\text{H}}^4}{3z_2^2}\left(1+\frac{m-1}{N}\right)^{-1}\frac{2}{N^2}\left(am+b-c\sqrt{m}\right),
\end{equation}
with fit parameters $a,\,b,\,c$ that inherit the $N$-dependence.
Note that we are now considering the normalized Fisher information again.

The fit function of the Fisher information in Eq.~\eqref{eq app: fit Fisher information} is fitted over $m\in\{2,3,\dots,20\}$ to the numerical results calculated via Eq.~\eqref{eq app: superradiance numerical integral} for every $N\in\{2,3,\dots,20\}$ independently.
Fig.~\ref{fig app: fit parameters a(N), b(N), c(N)} shows all fit parameters depending on $N$.
At first glance, $a(N)$ and $b(N)$ coincide to a certain extent, and $c(N)$ seems to be twice as large.
Furthermore, all curves indicate a power-law dependence on $N$.
\begin{figure}[h!]
    \includegraphics[scale = 0.9]{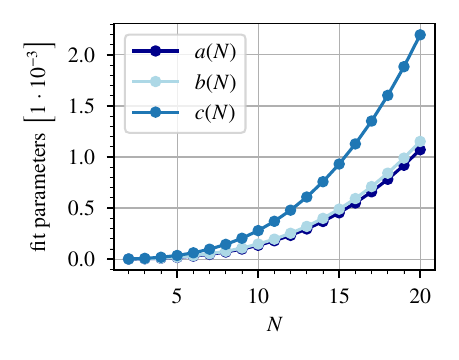}
    \caption{The introduced fit parameters $a,\,b,\,c$ of Eq.~\eqref{eq app: fit Fisher information} as a function of the number of thermal light sources $N\in\{2,3,\dots,20\}$ when fitted over $m\in\{2,3,\dots,20\}$ to Superradiant LIDAR's numerically evaluated Fisher information.
    A power-law dependence is evident.}
    \label{fig app: fit parameters a(N), b(N), c(N)}
\end{figure}
Both of the aforementioned observations are to be expected as $c=2a=2b=0.25N^3$ reproduces the previously discussed lower-bound solution of the Fisher information.

In order to reduce the number of fit parameters and analyze the $N$-dependence of the displayed parameters $a(N),\,b(N)\,\text{and}\,c(N)$ more thoroughly, we perform a subsequent fit with a power-law function defined as
\begin{equation}
    \label{eq app: power-law fit}
    f(N) = pN^e.
\end{equation}
Here, the prefactor $p$ and exponent $e$ are the new fit parameters.
This fit is performed for all parameters $a(N),\,b(N)\,\text{and}\,c(N)$ respectively and yields the six values displayed in Tab.~\ref{tab app: fit parameters}.
Evidently, all exponents are approximately equal to three, which supports the assumption of a linear $N$-dependence made for the above-discussed general lower-bound formula of the Fisher information.
Furthermore, the prefactor for $c(N)$ is roughly twice as large as that for $a(N)$ and $b(N)$, which is assumed for the lower-bound solution as well.
However, instead of $0.25$, the prefactor for $c(N)$ equals approximately $0.3$.
\begin{table}[h!]
    \centering
    \begin{tabular}{c|c|c}
         & $p$ & $e$ \\ \hline\hline
        $a(N)$           & $0.140$ & $2.986$ \\ \hline
        $b(N)$           & $0.160$ & $2.965$ \\ \hline
        $c(N)$           & $0.294$ & $2.977$
        \end{tabular}
    \caption{Prefactors $p$ and exponents $e$ for fitting the power-law function in Eq.~\eqref{eq app: power-law fit} to the factors $a(N),\,b(N),$ and~$c(N)$. All exponents are approximately equal to 3, and the prefactor for $c(N)$ is approximately twice as large as that for $a(N)$ and $b(N)$.}
    \label{tab app: fit parameters}
\end{table}

\begin{figure}[h!]
    \begin{minipage}{0.48\textwidth}
        \includegraphics[scale = 0.9]{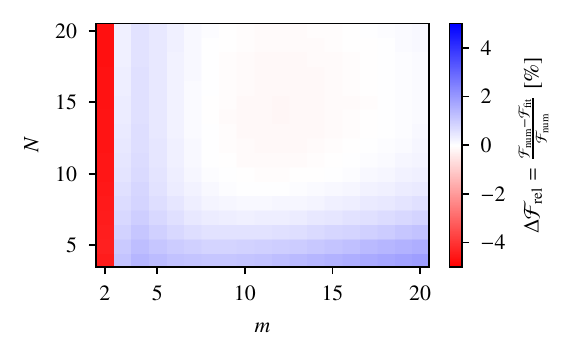}
    \end{minipage}
    \begin{minipage}{0.48\textwidth}
        \includegraphics[scale = 0.9]{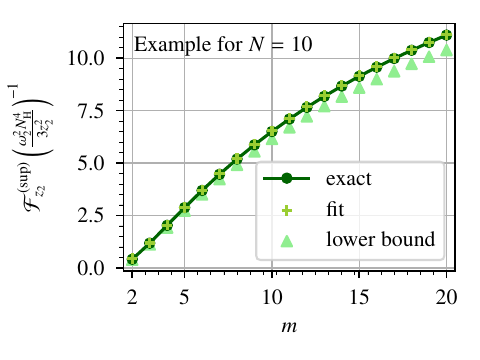}
    \end{minipage}
    \caption{Left: The relative difference $\Delta\mathcal{F}_\mathrm{rel}$ between the numerically evaluated Fisher information of Superradiant LIDAR (Eq.~\eqref{eq app: superradiance numerical integral}) and the Fisher information obtained via the fit procedure (Eqs.~\eqref{eq app: fit Fisher information} and~\eqref{eq app: power-law fit}) displayed for $N\in\{4,5,\dots,20\}$ and $m\in\{2,3,\dots,20\}$.
    The cases $N=2,3$ are not shown, as analytical solutions are presented in the main text. Negative differences indicate an overestimation of the Fisher information. All absolute values of the relative differences are below 5\,\%. Right: The Fisher information of Superradiant LIDAR for $N=10$ thermal light sources and $m\in\{2,3,\dots,20\}$. The numerically determined Fisher information (dots) is compared to the Fisher information determined via the lower-bound formula (triangles) shown in Eq.~\eqref{eq app: unnormalized lower bound Fisher information}, and the fit procedure (crosses). The fit procedure describes the Fisher information more accurately, whereas the lower-bound solution always underestimates it.}
    \label{fig app: fit discussion}
\end{figure}

We evaluate the accuracy of the performed fits by calculating and plotting the relative differences $\Delta\mathcal{F}_\mathrm{rel} = \left(\mathcal{F}_\mathrm{num} - \mathcal{F}_\mathrm{fit}\right)/\mathcal{F}_\mathrm{num}$ of the numerical results and the Fisher information determined by the fits of Eqs.~\eqref{eq app: fit Fisher information} and~\eqref{eq app: power-law fit}, similar to the validation of the lower-bound solution in the main text.
These relative differences are shown in the left panel of Fig.~\ref{fig app: fit discussion}.
Again, we do not display the relative differences for $N=2,3$ light sources, as analytical solutions for these cases are presented in the main text.
As can be seen, the relative differences are not solely positive.
In particular, the differences for correlation order two stand out as they are negative and show the largest deviation of approximately 4.5\,\%.
All further negative deviations are below 0.2\,\%.
Although most negative deviations are very small, they show that the fit procedure overestimates the Superradiant LIDAR's Fisher information for some configurations. 
This has to be kept in mind for the interpretation of the Cramér-Rao bound because its overestimation is not guaranteed.
Still, the accuracy of this fit procedure is indeed better than that of the lower-bound expression of the Fisher information (compare with Fig.~\ref{fig app: lower bound discussion}).
This can also be seen by comparing the Fisher information itself for all procedures.
The right panel of Fig.~\ref{fig app: fit discussion} shows this comparison for $N=10$ thermal light sources as an example.
It is clearly visible that the lower-bound solution always underestimates the exact Fisher information.
Evidently, using the fit results to describe the Fisher information yields considerably more accurate results.
However, it overestimates the Fisher information for $m\in\{2,10,11,12,13,14\}$.
Still, it is important to note that all the overestimations, except for $m=2$, are below 0.06\,\%.

To summarize the discussion, we have motivated how to find an approximate expression for the general Fisher information that always underestimates the exact results, but describes the former reasonably well. To have a more accurate approximation, we introduced a fit function that justifies the physically motivated lower-bound solution to a certain extent.
Namely, it supports the linear increase in $N$ of the Fisher information that is assumed for the lower bound.
Although the fit procedure is more accurate, we note that a more careful interpretation has to be done, since the fit result overestimates the Fisher information for certain configurations.
Finally, we note that if one wants to analyze the functional behavior of the Fisher information for different intervals of $N$ and $m$ than discussed in this manuscript, a rough approximation is always given by the lower-bound solution, whereas a fit with the introduced fit functions delivers a more accurate result.


\end{document}